\documentclass[fleqn, usenatbib]{mnras}
\usepackage[fleqn]{amsmath}
\usepackage{amssymb}
\usepackage[none]{hyphenat}
\usepackage{times}
\usepackage{graphicx}
\usepackage{color}
\usepackage{blindtext}
\usepackage{gensymb}
\title[Oort cloud asteroids]{Oort cloud asteroids: Collisional evolution, the Nice Model, and the Grand Tack.}

\author[Shannon et al.]{Andrew Shannon$^{1,2,3}$\thanks{E-mail: abshannon@psu.edu}, Alan P. Jackson$^{4,5}$, Mark C. Wyatt$^{3}$\\%
$^1$Department of Astronomy and Astrophysics, The Pennsylvania State University, 525 Davey Laboratory, University Park, Pennsylvania 16802, USA\\%
$^{2}$Center for Exoplanets and Habitable Worlds, The Pennsylvania State University, State College, Pennsylvania, USA \\%
$^3$Institute of Astronomy, University of Cambridge, Madingley Road, Cambridge, CB3 0HA, UK \\%
$^4$Centre for Planetary Sciences, University of Toronto, 1265 Military Trail, Toronto, Ontario, M1C 1A4, Canada\\%
$^5$School of Earth and Space Exploration, Arizona State University, 781 E Terrace Mall, Tempe, Arizona 85287, USA}
\date{submitted 2018}
\pagerange{\pageref{firstpage}--\pageref{lastpage}}
\pubyear{2018}

\begin{document}
\label{firstpage}

\maketitle

\begin{abstract}

If the Solar system had a history of planet migration, the signature of that migration may be imprinted on the populations of asteroids and comets that were scattered in the planets' wake.  Here, we consider the dynamical and collisional evolution of inner Solar system asteroids which join the Oort cloud.  We compare the Oort cloud asteroid populations produced by migration scenarios based on the `Nice' and `Grand Tack' scenarios, as well as a null hypothesis where the planets have not migrated, to the detection of one such object, C/2014 S3 (PANSTARRS).   Our simulations find that the discovery of C/2014 S3 (PANSTARRS) only has a $> 1\%$~chance of occurring if the Oort cloud asteroids evolved on to Oort cloud orbits when the Solar system was $\lesssim 1 \rm{Myr}$~old, as this early transfer to the Oort cloud is necessary to keep the amount of collisional evolution low.  We argue this only occurs when a giant $\left(\gtrsim 30 m_{\oplus}\right)$~planet orbits at $1\sim 2 \rm{au}$, and thus our results strongly favour a `Grand Tack'-like migration having occurred early in the Solar system's history.
  
\end{abstract}


\section{Introduction}
\label{sec:intro}

The Oort cloud is a population of small bodies, predominantly comets, found today at $10^4 \sim 10^5$~au from the Sun, but formed in the inner $\sim 35$~au~of the solar system \citep{1950BAN....11...91O,2008ssbn.book..315D}.  After formation, they were scattered outwards by the planets, and lodged in the outer solar system by perturbations from passing stars, giant molecular clouds, and the galactic tide \citep{1986Icar...65...13H,1987AJ.....94.1330D,2004ASPC..323..371D}.  Although most of these objects are expected to have been produced beyond Saturn \citep{1981Icar...47..470F}, \citet{1997ApJ...488L.133W} first postulated that a small fraction should be produced in the inner Solar system, and have the characteristics of asteroids, not cometary nuclei.  The first unambiguous Oort cloud asteroid, C/2014 S3 (PANSTARRS) was discovered by \citet{2016SciA....2E0038M}.  The total mass of Oort cloud comets has been suggested as a constraint on the proto-Solar nebula mass distribution \citep{2010A&A...509A..48P}, the dynamical history of the planets \citep{2008A&A...492..251B,2013Icar..225...40B}, and the cluster environment in which the Sun formed \citep{2010Sci...329..187L,2012Icar..217....1B}.  The semimajor axes of Oort cloud comets has been suggested for use as a constraint on the environment in which the Sun formed \citep{1997Icar..129..106F,2008Icar..197..221K,2011Icar..215..491K}.  In this tradition, \citet{2015MNRAS.446.2059S} suggested that measuring the fraction of Oort cloud objects which are asteroids, rather than comet nuclei, could provide an important constraint on the migration history of the Solar system's planets.  They provided an estimate that $\sim 4\%$~of Oort cloud objects should be asteroids, however, this estimate relied on the stated assumption that any dynamical evolution of the planets was unimportant, and the unstated assumption that any collisional evolution was unimportant. 

The orbits of the planets are generally believed to have evolved significantly over the history of the Solar system \citep{2018ARA&A..56..137N}.  \citet{1984Icar...58..109F,1996P&SS...44..431F} showed that scattering of residual planetesimals would drive an outward migration of Neptune, Uranus, and Saturn, and a corresponding inward migration of Jupiter.  \citet{1993Natur.365..819M,1995AJ....110..420M} found that this migration explained the mean motion resonance between Neptune and (134340) Pluto and would predict other such resonant objects \citep{1999AJ....117.3041H,2005AJ....130.2392H}, which was quickly confirmed \citep{1995AJ....109.1867J,1995Icar..118..322M}.  The dynamical path of this migration can be constrained by the populations of resonant objects \citep{2002AJ....124.3430C,2002MNRAS.336..520Z,2013CoSka..43..119N,2014AJ....148...56B}, and the preservation of the Cold Classical Kuiper belt \citep{2011ApJ...738...13B,2012ApJ...746..171W,2012ApJ...750...43D,2015AJ....150...68N}.  The timing of this migration is often associated with the Late Heavy Bombardment \citep{2005Natur.435..466G,2005Natur.435..459T,2008Icar..196..258L,2011AJ....142..152L}, several hundred million years after the birth of the Solar system, however this association is not universal \citep[e.g.,][]{2012ApJ...745..143A,2018Icar..311..340C}.  Despite the possible constraints offered from observations, the enormous range of possible histories leaves this an area of active research \citep[e.g.][]{2015MNRAS.451.2399C,2015AJ....150...73N,2015Icar..247..112P,2015ApJ...806..143V}, where not even the number of participating planets is known \citep{2006ApJ...643L.135G,2008AJ....135.1161L,2009Icar..204..330Y,2011ApJ...742L..22N,2012AJ....144..117N,2012ApJ...744L...3B}.  Apart from perhaps the loss of Theia at $10^7 \sim 10^8 \rm{years}$~\citep{1973LPI.....4..723T,1976LPI.....7..120C,2015Icar..248..318Q}, the terrestrial planets are often thought of as not having evolved significantly, although this is far from certain \citep{2007Icar..189..386C,2015ApJ...806L..26V}.

Models of the collisional evolution of comets during the Solar system's evolution suggest that collisional evolution may have played a significant role in setting the present day population \citep{2001Natur.409..589S,2003Icar..166..141C,2007Icar..188..468C,2017A&A...597A..61J}.  Observations of comet 67P/Churyumov-Gerasimenko, however, suggests it may not have undergone significant collisional evolution \citep{2015Natur.526..402M,2016A&A...592A..63D}, and \emph{New Horizons}~data reveals a lack of small craters on Pluto and Charon that may be indicative of little collisional evolution in the outer Solar system \citep{2016LPI....47.2310S}.  In the inner Solar system, it is clear that collisional evolution cannot be neglected \citep{2005Icar..179...63B,2015aste.book..701B,2016E&PSL.455...85B}.  The collisional evolution of a population of bodies which is also dynamically evolving is a computationally hard problem.   Various codes have been developed to couple the dynamical and collisional evolution, such as \citet{2012AJ....144..119L,2013A&A...558A.121K,2015A&A...573A..39K,2013ApJ...777..144N,2015ApJ...798...83N}, however the computation demands of this approach have limited the range of problems that it can be applied to.

In this work, we consider the implications of the existence of the Oort Cloud Asteroid C/2014 S3 (PANSTARRS) for the dynamical history of the Solar system.  We describe how we model the dynamical and collision evolution of the Solar system for different dynamical histories in \textsection \ref{sec:method}.  We present the results of our models, and calculate how likely they are to produce C/2014 S3 (PANSTARRS)-like objects in \textsection \ref{sec:results}.  We discuss the implications of this in \textsection \ref{sec:discussion}. 

\section{Method}
\label{sec:method}

We wish to model both the dynamical and collisional evolution of a population of minor planets.  To make the problem more computationally tractable, we split the modelling into two steps.  First, we perform $N$-body simulations of the dynamical evolution of the Solar system, including test particles to represent the minor planets (\textsection \ref{subsec:dynamicsmethod}), then we post process the results of those simulations to collisionally evolve the minor planets by mutual collisions (\textsection \ref{subsec:collisionsmethod}).  Because catastrophic impacts are dominated by the smallest possible impactors, which are typically an order of magnitude or more smaller in mass than the body disrupted, we do not expect significant collisional damping, allowing this separation.

\subsection{Dynamics}
\label{subsec:dynamicsmethod}
To calculate the dynamical evolution of the minor planets, we performed $N$-body simulations with \textsc{Mercury} \citep{1999MNRAS.304..793C}, modified to account for the Galactic tide per \citet{2013MNRAS.430..403V} and stellar flybys per \citet{2014MNRAS.445.4175V}, necessary to account for the formation of Oort cloud objects.  We use test particles to represent the minor planets.  The total mass expected to reside in the minor planets is sufficient to be dynamically important, however, as we are interested in the consequences of specific dynamical histories, we apply additional forces to the planets to produce the migration histories we desire.  As in \citet{2015MNRAS.446.2059S}, we perform separate simulations for batches of test particles with initial semimajor axis in 1 au bins, \textcolor{black}{except the innermost bin, which extends from 0.5 au to 1 au.  We use} 1000 particles per au, except where long term stability in the asteroid belt, or beyond 30 au, makes this computationally infeasible.  In that case, we use 100 particles per au.  Test particles begin with low eccentricity $\left( 0 \leq e \leq 0.05\right)$, \textcolor{black}{low inclination $\left(0 \degree \leq i \leq 4.5 \degree\right)$~orbits.}
For initial orbits below 6 au, we include all eight planets and use an eight day timestep.  Beyond that, we include only the giant planets, and use a 120 day timestep.  Particles were removed once they reached $250,000$~au from the Sun.  Simulations were run for $4.5$~Gyr, or until all particles were lost due to collisions with the Sun or a planet, or were ejected.  We label particles that begin with $a < 3$~au as `asteroids' and particles beyond that as `comets', to distinguish objects that are likely to be volatile-free and so do not outgas from those likely to be volatile-rich that do.  \textcolor{black}{The representative ice line of the Solar system is typically placed at this location \citep[e.g.,][]{2007AsBio...7...66R,2012M&PS...47.1941W}, based on Solar system evidence \citep{2012AREPS..40..251M,2014Natur.505..629D} with various possible theoretical underpinnings \citep[e.g.,][]{2012MNRAS.425L...6M,2016Icar..267..368M}}

We consider four possible dynamical histories for the Solar system.  For the static Solar system case, we begin with the planets on their orbits as of epoch 2451000.5, and apply no evolution to them other than their mutual gravity.  This is essentially the same setup used in \citet{2015MNRAS.446.2059S}, but we perform new simulations as the simulations in that work were performed with a version of \textsc{Mercury}~which was not patched for the known bug that causes erroneous detections of collisions with the central body \citep{2013AsBio..13.1155W} \footnote{Thanks for B. Gladman for calling this to our attention}.

\begin{figure}
\includegraphics[width=0.33\textwidth, angle=270, trim = 0 0 0 0, clip]{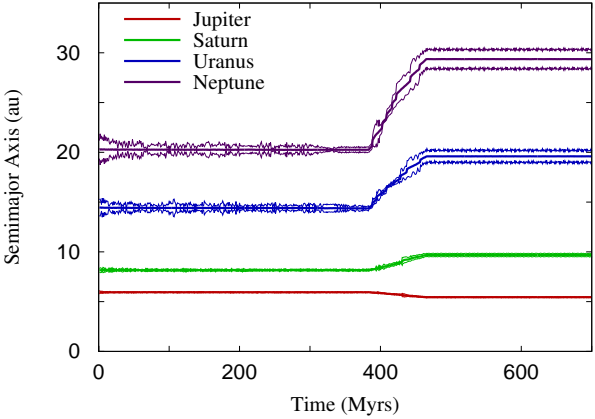}
  \caption{Semimajor axes (thick lines) and perihelion and aphelion (thin lines) of Jupiter (red), Saturn (green), Uranus (blue), and Neptune (purple) in our Nice-like scenario.  The planets begin with a more compact configuration than the current Solar system, and begin to expand after $\sim 380$~Myrs, arriving at their current orbits after $\sim 80$~Myrs.  \textcolor{black}{The current understanding is that this migration should have taken $10\sim 100$~Myrs \citep{2015AJ....150...73N}}.}
  \label{fig:nicedynamics}
\end{figure}

The second dynamical history we consider is inspired by the Nice model \citep{2005Natur.435..466G,2005Natur.435..462M,2005Natur.435..459T}.  The Nice model has undergone numerous expansions and modifications since its inception \citep[e.g.][]{2007AJ....134.1790M,2008Icar..196..258L,2012Natur.485...78B,2013ApJ...768...45N}, so we focus on the core premise of the orbital expansion of the outer Solar system giant planets \citep{1984Icar...58..109F,1993Natur.365..819M} occurring when the Solar system was several hundred million years old.  The timing of the migration, and the exact dynamical path taken by the giant planets remains a subject of inquiry \citep[e.g.,][]{2012ApJ...750...43D,2014Icar..232...81M,2015AJ....150..157B,2015AJ....150...73N,2015Icar..247..112P,2016AJ....152..133K,2018Icar..306..319G}.  We produce a qualitatively similar history by applying additional forces to the giant planets after a few hundred million years to produce the desired migration, as well as to damp the eccentricities.  The eccentricity damping produces a somewhat less violent evolution than is often assumed for the Nice model, but allows us to avoid wasting computation time on runs that eject a giant planet \citep[as in e.g.,][]{2011ApJ...742L..22N,2012AJ....144..117N}.  The migration is plotted in figure \ref{fig:nicedynamics}.

\begin{figure}
\includegraphics[width=0.33\textwidth, angle=270, trim = 0 0 0 0, clip]{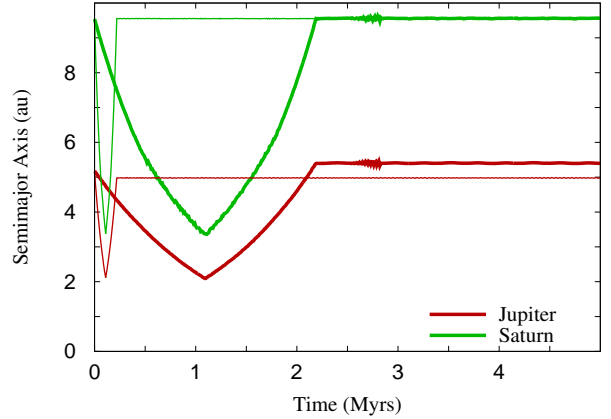}
  \caption{Semimajor axis evolution of Jupiter (red) and Saturn (green) in our Fast Tack (thin lines) and Slow Tack (thick lines) scenarios.  The planets migrate inwards until Jupiter hits $\sim 2$~au, when the migration reverses and the planets return to their present-day orbits.}
  \label{fig:tackdynamics}
\end{figure}

The third and fourth dynamical histories we consider are inspired by the Grand Tack model \citep{2011Natur.475..206W,2012M&PS...47.1941W}.  In the Grand Tack, Jupiter acquires roughly its current mass before Saturn does, and undergoes type II migration.  Saturn then acquires (roughly) its current mass, and undergoes type II migration at a faster rate than Jupiter.  As Saturn approaches Jupiter, the planets become caught in a mean-motion resonance, and the direction of migration reverses, with Jupiter and Saturn arriving at their starting location for the later expansion of the outer planets as the gas disk dissipates.  We consider two variants, one where the migration timescale is $\sim 10^5$~years, which we label `Fast Tack', and one where the migration timescale is $\sim 10^6$~years, which we label `Slow Tack'\textcolor{black}{, motivated by the result that standard Type II migration may be unable to reproduce the observed exoplanet population, and a variety of physical mechanisms may be able to slow it \citep{2013ApJ...774..146H}}.  The orbital evolution of Jupiter and Saturn in both Tack cases are plotted in figure \ref{fig:tackdynamics}.  In both Tack cases, we do not simulate test particles beyond 14 au, but instead use the particle tracks from the Static case for those bodies.  As neither Jupiter nor Saturn scatter these bodies directly, and as in both of our Tack cases Jupiter and Saturn return to roughly their current orbits well before Uranus or Neptune can effectively scatter bodies away \citep{2015ApJ...799...41M,2016MNRAS.462L.116S}, we do not expect these bodies to undergo a significantly different dynamical evolution.

\subsection{Collisions}
\label{subsec:collisionsmethod}

Consider an individual minor planet.  At any given location the number density of minor planets is $n$ and their mean relative velocity with respect to the minor planet we are following is $v_{\rm rel}$.  If the minor planet we are following has a cross-sectional area for collision, $\sigma$ (which may be larger than its physical cross-sectional area due to gravitational focussing), then at that location the rate at which our minor planet will experience collisions is
\begin{equation}
    r_{\rm col} = n \sigma v_{\rm rel}.
\end{equation}
It is possible that $n$ and $v_{\rm rel}$ will vary along the orbit of our minor planet, especially if its orbit is eccentric or inclined.  If gravitational focussing is important then $\sigma$ may also vary since it is then dependent on $v_{\rm rel}$.  We must then average the local collision collision rate, $r_{\rm col}$, around the orbit of our minor planet to find the orbit-averaged collision rate
\begin{equation}
    R_{\rm col} = <r_{\rm col}> = <n \sigma v_{\rm rel}>,
\end{equation}
where the angle brackets indicate an average around the orbit.  Minor planets on different orbits will experience different values of $n$ and $v_{\rm rel}$ and thus $R_{\rm col}$ even if they have the same size.  It is these collision rates that are key to computing the collisional evolution of the minor planet population.

To determine the collision rates and compute the collisional evolution of our bodies we use the collisional evolution code of \citet{jackson2012, jackson2014}.  Since the number of particles that we can include in our $N$-body simulations is necessarily much smaller than the expected number of minor planets we do not assume that each particle is representative of an individual minor planet.  Instead the collisional code assumes that each of the $N$-body test particles is representative of a population of bodies on similar orbits and that this population of bodies is described by a power-law \textcolor{black}{differential} size-distribution in the diameter, $D$, such that, $n(D)dD \propto D^{-q}dD$, and a maximum size $D_{\rm max}$.  The slope of the size-distribution, $q$, and $D_{\rm max}$ are assumed to be the same for all of the $N$-body particles, but the initial mass (and thus the constant of proportionality) can be set individually.

Note that strictly the local $v_{\rm rel}$ can also be a distribution rather than a single value, since the orbit of our minor planet can be intersected by a range of different orbits at any given location.  There is however a trade-off, computing the local $v_{\rm rel}$ distribution would require taking a substantial number of nearest neighbours, which would increase how computationally intensive the calculation is as well as enlarging the nearest neighbour volume, which can degrade the accuracy, especially in low density regions.  As such our code makes the choice of using a smaller number of nearest neighbours (10 as standard) and using the mean value.

We employ $D_{\rm max} = 120 {\rm km}$, the primordial peak of the mass distribution observed among asteroids \citep{2005Icar..175..111B,2009Icar..204..558M}, and other small body populations in the Solar system \citep{2010ApJ...723L.233S,2014ApJ...782..100F}.  Current theories of planetesimal formation and growth produce top-heavy mass distributions of planetesimals \citep{2015SciA....1E0109J,2016ApJ...818..175S,2017ApJ...847L..12S}, although the size of the largest planetesimals is not yet robustly predicted.  \textcolor{black}{Although there are larger objects in the Solar system, $D_{\rm max} = 120 {\rm km}$~represents the peak of the mass normalisation of the collisional cascade \citep[e.g.,][]{2008ApJ...673.1123L,2011ApJ...739...36S} - even at sizes that have since been lost \citep{2018MNRAS.480.1870S}, and is thus the most representative choice for the collisional model.  Thus, we do not include any such larger objects in our collisional cascades, but their inclusion would only slightly alter the evolution, and only at very late times.}  We set $q = 3$~as this value is appropriate for \textcolor{black}{collisionally evolved} large bodies bound by gravity \citep{1997Icar..130..140D}.  \textcolor{black}{Observational \citep[e.g.,][]{2017Icar..287..187R,2019arXiv190210795S} and theoretical results \citep[e.g.,][]{2017ApJ...847L..12S} may also favour it as the primordial size-number distribution.} The normalisation of planetesimal masses is set such that the surface density at 1 au is $17 {\rm g/cm^2}$~and falls as $\Sigma \propto a^{-3/2}$, giving it the characteristics of the Minimum Mass Solar Nebula (MMSN) \textcolor{black}{\citep{1977Ap&SS..51..153W,1981PThPS..70...35H,2014A&A...570A..35M}.  As all our simulations will undergo multiple e-foldings of collisional depletion in this region, the exact choice will not affect the outcome, unless it is more than an order of magnitude too high \citep{2007ApJ...658..569W}}.  Beyond 3 au, we double the surface mass density to account for the contribution of ice.  We do not model changes in the slope of the size distribution or the maximum size and thus assume that this distribution with $q=3$ is also the steady-state distribution.  Utilising a $q=3$ distribution required a minor modification to the code of \citet{jackson2014}, since the mass integral must be performed differently to avoid a singularity in this case.

Now that we have our size-distribution we can compute the collision rates.  As we described above we assume that each of our $N$-body test particles is representative of a whole population of bodies on similar orbits, and need to average the local collision rates to compute the orbit-averaged collision rate.  To compute the orbit-averaged rate we re-sample the $N$-body particles and randomise the mean anomaly, which assumes that there is nothing special about the location of the $N$-body particle on its orbit at any given output from the $N$-body simulation.  In addition, since the number of $N$-body particles is fairly small we also randomise their arguments of pericentre and longitudes of ascending node.  The argument of pericentre and longitude of ascending node vary on timescales much shorter than the age of the Solar system due to precession and so in randomising these angles we assume that each $N$-body test particle is representative of one snapshot in the precession cycle of a population of objects with the same semimajor axis, eccentricity and inclination and that there is nothing special about this snapshot.  Each $N$-body particle thus becomes a toroid of collisional particles.  In total we use $10^5$~collisional particles for our calculations, resampling each $N$-body particle roughly $3$~times.  \textcolor{black}{This randomisation of these orbital angles has long been used in calculating collisional probabilities \citep{1951PRIA...54..165O}, and reasonably reproduces more complicated calculations \citep{1999Icar..142..509D}, except in a few pathological cases (such as when a large fraction of particles are in a single resonance) which we do not expect to be significant here.  Use of this assumption thus means we are implicitly assuming that resonant objects make a negligible contribution to the population.  Employing a sufficient number of particles to make the random angle approximation unnecessary would be computational prohibitive for this project, however since we do not expect resonant objects to account for a significant fraction of the population use of the random angle approximation should not significantly affect our results.}

The collision rates calculated at each $N$-body output timestep then allow us to compute the evolution of the collisional cascade over time.  One advantage of our $q=3$ size-distribution, and indeed any size-distribution with a power-law slope shallower than $q=4$, is that the mass of the cascade is dominated by the largest objects in the distribution so that it is the break-up of these objects that drives the evolution of the whole cascade and it is their lifetime that sets the evolution timescale of the whole distribution.  The lifetime of the largest objects is dependent on the number of objects capable of colliding with them catastrophically, and so on $D_{\rm cc}(D_{\rm max})$, the size of the smallest object that can catastrophically collide with an object of size $D_{\rm max}$.  Our collisional code calculates $D_{\rm cc}(D_{\rm max})$ using the velocity dependent dispersal threshold of \citet{stewart2009}.  The rate at which the largest objects experience catastrophic collisions is then $R_{\rm cc} = < n \sigma_{\rm cc}(D_{\rm max}) v_{\rm rel} >$, where $\sigma_{\rm cc}(D_{\rm max})$ is the cross-section for catastrophic collisions for an object of size $D_{\rm max}$ and is given by
\begin{equation}
    \sigma_{\rm cc} = \int_{D_{\rm cc}(D_{\rm max})}^{D_{\rm max}} n(D) \left( \frac{D_{\rm max} + D}{2} \right)^2 dD,
    \label{eq:sigmacc}
\end{equation}
neglecting gravitational focussing.  The lifetime of the largest objects, $\tau$, is then $R_{\rm cc}^{-1}$ and the mass of the cascade evolves as
\begin{equation}
    m(t + \delta t) = m(t) \frac{1}{1 + \delta t / \tau},
    \label{eq:massevol}
\end{equation}
where $\delta t$ is the time between $N$-body outputs.  The collision rates, and $\tau$ are re-calculated at each $N$-body output, but note that Equation~\ref{eq:massevol} treats $\tau$ as a constant.  This is reasonable as long as the variation in $\tau$ between $N$-body outputs is small.  The lifetime of the largest objects, and the mass evolution is calculated separately for each $N$-body particle, allowing the evolution to proceed at different rates in different parts of the Solar system.

\section{Results}
\label{sec:results}

\subsection{Test Particle Dynamics}
\label{subsec:dynamicsresults}

\begin{figure}
\includegraphics[width=0.48\textwidth, angle=0, trim = 0 0 0 0, clip]{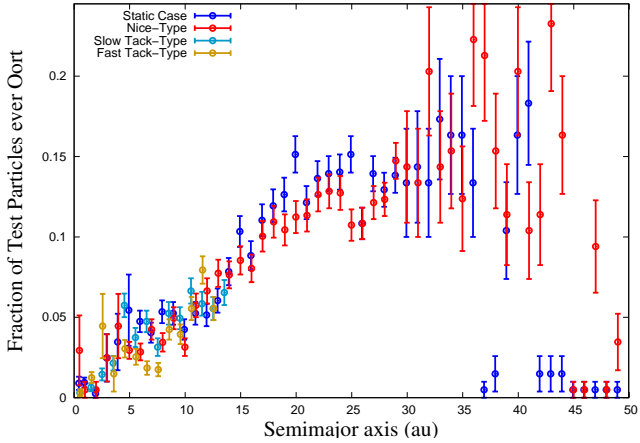}
  \caption{Fraction of test particles that become Oort cloud objects as a function of initial semimajor axis, in our Static case (blue), Nice case (red), Slow Tack case (turquoise), and Fast Tack case (yellow).  The fraction of bodies that become Oort cloud objects is roughly the same in all four dynamical histories.  The uncertainties plotted here are Wilson intervals \citep{10.2307/2276774}.
  }
  \label{fig:oortfrac}
\end{figure}

To define membership in the Oort cloud, we label all particles members of the Oort cloud once they have semimajor axis $a > 1000 \rm{au}$~and perihelion $q > 50 \rm{au}$.  \textcolor{black}{The former limit allows us to capture all the objects interacting with the galactic potential and passing stars while excluding everything in or near the original disk \citep{2004ASPC..323..371D}, while the latter excludes scattered disk objects interacting with Neptune \citep{2002Icar..157..269G}.}  In figure \ref{fig:oortfrac}, we plot the fraction of test particles that are ever members of the Oort cloud as a function of their initial semimajor axes, comparing the results from the different dynamical histories presented in section \ref{subsec:dynamicsmethod}.  It is critical to note that this figure presents only the test particle dynamics, and does not include the depletion of particles due to collisional evolution.  In all four dynamical histories, the fraction of test particles that become members of the Oort cloud is roughly the same at any semimajor axis.  The fraction of test particles that ever become Oort cloud objects increases with their primordial semimajor axis, which is expected as the time needed for a planet to eject a test particle increases with semimajor axis, making it more likely the Galactic tide can capture the test particle into the Oort cloud \citep{1993ASPC...36..335T}.

\subsection{Collisional Evolution}
\label{subsec:collisionsresults}

\begin{figure}
\includegraphics[width=0.33\textwidth, angle=270, trim = 0 0 0 0, clip]{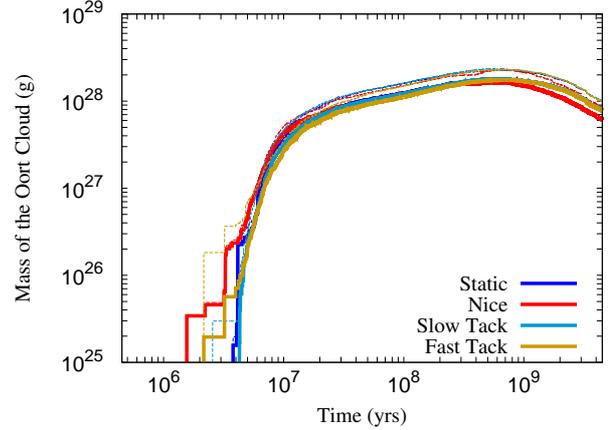}
  \caption{Mass evolution of the entire Oort cloud in the four histories that we consider, with Static in blue, Nice in red, Slow Tack in turquoise, and Fast Tack in yellow.  The thin, dashed lines do not include collisional evolution, while the thick, solid lines do include collisional evolution.  All four cases produce similar Oort cloud clouds, with total masses of roughly one Earth mass, compatible with the observed total mass of the Oort cloud today.  Collisional evolution does not significantly impact the overall evolution of the cloud.}
  \label{fig:wholeoortmass}
\end{figure}

As a viability test of our models, we plot the mass evolution of the entire Oort cloud within our simulations in figure \ref{fig:wholeoortmass}.  The present day mass in all cases is $\sim M_\oplus$, similar to other dynamical models, and consistent with (though towards the low end of) observational estimates \citep{2008ssbn.book..315D}.  Collisional evolution does not drastically affect the overall evolution of the Oort cloud, reducing the overall mass by $20\% \sim 30\%$.  Because the objects do not experience significant collisional loss, the total mass could be increased somewhat by assuming a primordial disk that was a few times the level of the MMSN.  \textcolor{black}{We compare these results to the work of}  \citet{2007Icar..188..468C}, \textcolor{black}{who considered the collisional evolution of objects that would later form the Oort cloud.  The details of their implementation were significantly different from our own.  While we considered four different dynamical histories, they employed a single planet migration scheme based on \citet{1995AJ....110..420M}, where the orbits of the giant planets were more compact originally, but expanded to their current positions smoothly, starting from when the Solar system formed.  Their implementation of a size distribution and collisional evolution attached size bins to each particle, each of which then collisionally evolved under the \"{O}pik approximation \citep{1951PRIA...54..165O}, with fragments from destructive collisions added to smaller size bins attached to the same particle.  They considered a steeper initial size distribution with $q = 3.5$, which increased their rate of collisional evolution.  They employed a different collisional strength law, taking $Q^*$~from \citet{1999Icar..142....5B}, which for typical collisional velocities in these simulations would be $1\sim 3 \times$~stronger, and would also produce a $q \approx 3.5$ power law of collisional fragments.  \citet{2007Icar..188..468C} considered a surface density $\Sigma \propto a^{-1}$, which reduces the rate of collisional evolution, that the primordial disk extended from 5 \rm{au} to 50 \rm{au}~which left a massive belt to collide against at 40-50 \rm{au} that increased collisional evolution.  They considered cases where the primordial mass distribution peaked at sizes from 1 m to 100 km.  Comparing to their 100 km case (the most similar to our simulations), they found a $\sim 60\%$~mass reduction of the Oort cloud due to collisional dynamics.  That we are in good agreement despite all these different choices is a strong indicator the exact choices made here are not critically important. As their choice of size-number distribution increased their rate of collisional evolution while their other choices decreased it, this is likely the main cause behind the slight difference in our outcomes.  The choice of largest size is quite important, as in the most extreme case they find a mass reduction of $> 95\%$~in the case where the mass peaks at 100 m.  As we are able to well motivate the choice of initial largest size from Solar system observations, we do not consider such unphysical cases.}

\begin{figure}
\includegraphics[width=0.33\textwidth, angle=270, trim = 0 0 0 0, clip]{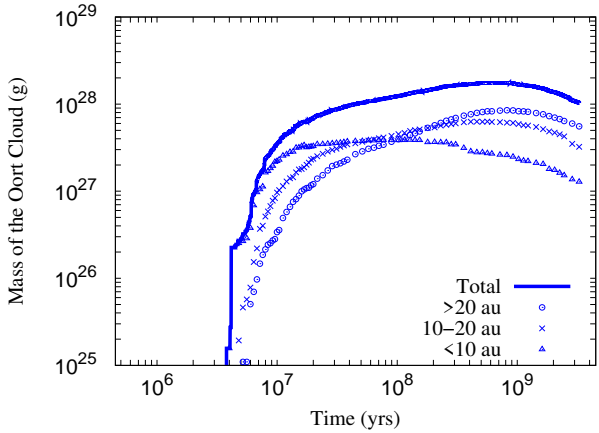}
  \caption{Origin region of the mass in the Oort cloud plotted as a function of time, using the example of our Static case.  The mass in Oort cloud objects formed at $\geq 20$~au is plotted with circles, the mass in Oort cloud objects formed at 10 au - 20 au is plotted with exes, and the mass in Oort cloud objects formed at $\leq 10$~au is plotted with triangles.  Most of the present day Oort cloud objects originated at $\geq 20$~au, as is expected given the icy, volatile-rich composition of typical Oort cloud objects.  The other cases (Nice, Fast Tack, Slow Tack) are very similar to the Static case in this regard.}
  \label{fig:cloudorigin}
\end{figure}

In figure 5, we plot the evolution of the mass of the Oort cloud, separating bodies originating at $< 10$~au, $10-20$~au, and $>20$~au to understand their relative contributions.  We find that the bulk of the present day mass in these models come from planetesimals that formed at $>20$~au (figure \ref{fig:cloudorigin}).  Thus, we would expect the Oort clouds generated in our simulations to be composed of icy bodies rich in volatiles, compatible with the observations that known Oort cloud objects are predominantly comets, often with a 'new' appearance.  This also accounts for why collisional evolution is only a mild effect, as at large semimajor axis surface densities are low, as are collisional velocities.

\begin{figure}
\includegraphics[width=0.33\textwidth, angle=270, trim = 0 0 0 0, clip]{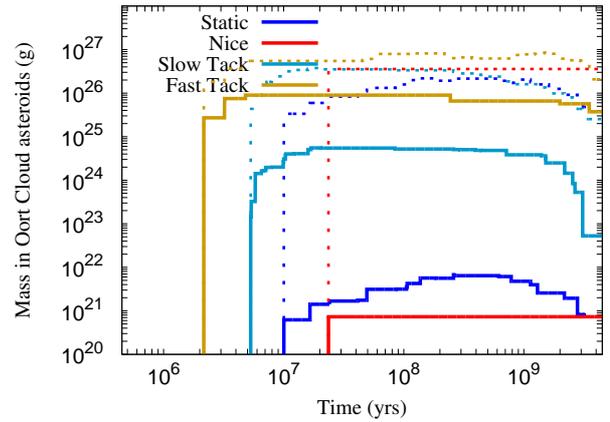}
  \caption{Evolution of the total mass in Oort cloud asteroids, considering collisional evolution (solid lines), compared to neglecting collisional evolution (dashed lines).  Here we present the results for four dynamical histories: the Static case (blue), where the giant planets have always had their present day orbits, the Nice case (red), where the original orbits of the giant planets were more compact, and they expanded to their present day orbits after $\sim 500$~Myrs, the Slow Tack case (turquoise), where Jupiter and Saturn migrate in to $\sim 2$~au, then return to their present orbits over $\sim 1$~Myr, and the Fast Tack case (orange), where Jupiter and Saturn migrate in to $\sim 2$~au, then return to their present orbits over $\sim 0.1$~Myr.  Neglecting collisional evolution, all dynamical histories produce similar evolutions, with their present day masses varying by less than an order of magnitude.  When collisional evolution is included, Grand Tack cases produce significantly more mass in Oort Cloud asteroids, as asteroids are transferred to the Oort cloud earlier, and thus undergo less collisional evolution.}
  \label{fig:evolvegrind}
\end{figure}

We focus now on the bodies we have labelled rocky asteroids, those that began with $a < 3$~au.    We plot the evolution of the total mass in Oort cloud asteroids in figure \ref{fig:evolvegrind}.  That plot contrasts how the total mass in Oort cloud asteroids evolves with time in all four dynamical histories, both with and without collisional evolution.  When collisional evolution is neglected, all four histories produce Oort clouds with a few percent of an Earth mass in asteroids, about $1\%$~of the test particles.  When collisional evolution is included, Grand Tack cases produce significantly more mass in Oort Cloud asteroids, as asteroids are transferred to the Oort cloud earlier, and thus undergo less collisional evolution.  In the Static and Nice cases, $99.99\% \sim 99.999\%$~of the asteroids are destroyed in catastrophic collisions before they join the Oort cloud, because those bodies reside in the inner Solar system, where they collisionally evolve for $10^7\sim 10^8$~years.  In contrast, in the Tack cases the asteroids that join the Oort cloud are sent there at $10^5\sim 10^6$~years, as they can be scattered to the Oort cloud directly by Jupiter and Saturn, rather than needing to rely on scattering by the terrestrial planets and resonances with the giant planets to raise their aphelia to large enough values for the asteroids to scatter off Jupiter.  This allows the asteroids to be scattered to the Oort cloud when only $90\% \sim 99\%$~of the asteroids has been lost to collisional destruction.

\begin{figure}
\includegraphics[width=0.33\textwidth, angle=270, trim = 0 0 0 0, clip]{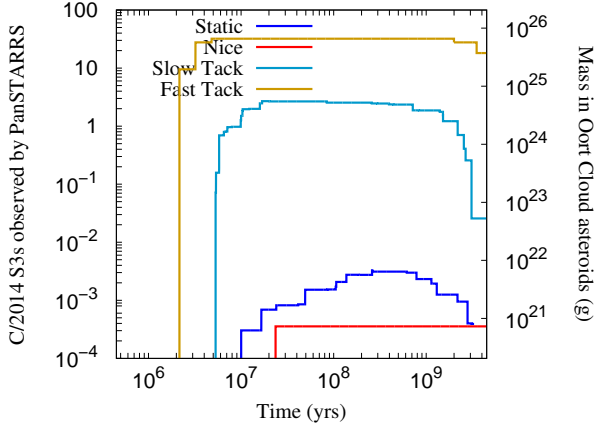}
  \caption{The number of Oort cloud asteroids as large as or larger than C/2014 S3 that would be detectable in the first ten tears of the operation of Pan-STARRS.  The right Y axis shows the corresponding total mass of Oort cloud asteroids.  Here we present the results for four dynamical histories: the Static case (blue), where the giant planets have always had their present day orbits, the Nice case (red), where the original orbits of the giant planets were more compact, and they expanded to their present day orbits after $\sim 500$~Myrs, the Slow Tack case (turquoise), where Jupiter and Saturn migrate in to $\sim 2$~au, then return to their present orbits over $\sim 1$~Myr, and the Fast Tack case (orange), where Jupiter and Saturn migrate in to $\sim 2$~au, then return to their present orbits over $\sim 0.1$~Myr.  The Tack cases can produce sufficient C/2014 S3-like objects to allow for their detection by Pan-STARRS, while non-Tack cases produce only a $\sim 0.1\%$~chance of any such detections.}
  \label{fig:panstarrs}
\end{figure}

To compare these predictions with observations, we consider C/2014 S3 (PANSTARRS), the first (and currently only) discovered rocky asteroid in the Oort cloud \citep{2016SciA....2E0038M}.  Assuming an albedo of $0.25$, typical of S-type asteroids, it has a radius of \textcolor{black}{$s \sim 0.25 \rm{km}$}.  Pan-STARRS can detect moving objects down to a magnitude of $\sim 21.7$~in a broad, visual filter \citep{2013PASP..125..357D}, allowing for the discovery of C/2014 S3-like objects at a maximum distance of $\sim 2.6 \rm{au}$~from the Sun.  To calculate the expected number of detectable C/2014 S3 analogues in our simulations, we do not assume an orbital distribution, instead, we return to our dynamical simulations in \textsection \ref{subsec:dynamicsresults}, and measure the instantaneous fraction of objects which had been labelled Oort cloud members that are within $2.6 \rm{au}$~of the Sun, which we find to be $2.5 \times 10^{-9}$. 
To convert from the mass of the Oort cloud to the number of objects with radius greater than or equal to \textcolor{black}{$0.25 \rm{km}$}, we calculate
\begin{equation}
  N_{>0.25 \rm{km}} = \frac{\int_{0.25 \rm{km}}^{60 \rm{km}} n_0 s^{-3} ds}{\int_{0\rm{km}}^{60\rm{km}}\frac{4\pi}{3}\rho s^3 n_0 s^{-3} ds} = 1.3 \times 10^{-17} \rm{g^{-1}},
  \label{eq:masstonumber}
\end{equation}

where $N_{>0.25 \rm{km}}$~is the number of objects with radius greater than $0.25$~km per unit mass, $n_0$~is a normalisation constant for the population, and we have assumed (as with the collisional model) that the population obeys $dn/ds \propto s^{-3}$~and has a maximum radius of $s = 60 \rm{km}$.  Following the general approach of \citet{2015MNRAS.446.2059S}, an object on an $ a \approx \infty, e \approx 1$~orbits, spends a time 

\begin{equation}
 \tau = \frac{4}{3}b^{1.5} \left(2GM_{\odot}\right)^{(-0.5)}
\end{equation}

within a distance $b$~of the Sun \textcolor{black}{\citep{1970Icar...13..231J}}, which is $\tau \sim 0.63$~years for $b = 2.6~\rm{au}$.   Pan-STARRS has operated for a decade, or $\sim 15$~passage times.  With these three quantities, we can convert the mass of the Oort cloud into an expected number of C/2014 S3 analogues Pan-STARRS would discover for a given history by converting the mass to a number using equation \ref{eq:masstonumber}, multiplying by the fraction that are detectable at any moment, then multiplying by the number of transit times Pan-STARRS was observing for.  We plot the results in figure \ref{fig:panstarrs}.   The Tack cases can produce sufficient C/2014 S3-like objects to allow for a significant chance of their detection by Pan-STARRS, while non-Tack cases produce only a $\sim 0.1\%$~chance of any such detections.

\section{Discussion}
\label{sec:discussion}

Considering four dynamical Solar system histories, we have shown the chance that available dynamical pathways will eventually lead asteroids to the Oort Cloud is largely independent of the dynamical history of the Solar system, but the time for the dynamical pathways to lead asteroids to the Oort cloud depends significantly on the assumed dynamical history.  As a consequence, the amount of collisional processing the asteroids undergo, and thus the number of asteroids that join the Oort cloud, depends strongly on the dynamical history of the Solar system.  We find that Grand Tack-like cases, where Jupiter and Saturn enter the inner Solar system at early times and directly scatter asteroids to the Oort Cloud can produce a sufficient number of Oort Cloud asteroids to make the discovery of C/2014 S3 (PANSTARRS) a likely event, while scenarios where the giant planets remain in the outer Solar system - our Static and Nice-like cases - produce only a $\sim 0.1\%$~of Pan-STARRS observing such an Oort Cloud asteroid.

Can we generalise this result?  Using the framework of \citet{2017MNRAS.464.3385W}, a giant planet $\left( m \gtrsim30 m_{\oplus}\right)$~at 1-2 au will directly eject asteroids, implanting a fraction in the Oort cloud.  If such a planet were in the inner Solar system when the Solar system was $\lesssim 1$~Myrs old, the necessary asteroids could be transferred to the Oort cloud before too much collisional evolution had taken place.  Without such a planet, the asteroids remain in the inner Solar system until the terrestrial planets can scatter them up to orbits where they interact with the giant planets, which takes sufficiently long that collisional evolution removes too much mass to allow the needed number of asteroids to persist.  Dynamical histories other than the Grand Tack that produce a giant planet in the inner Solar system at early times could also explain the existence of C/2014 S3 (PANSTARRS), but given that requirement, and the need to arrive at the present Solar system, it seems unlikely that the ultimate truth is not some variant of the Grand Tack.

Here we have considered the significance of a single object, C/2014 S3 (PANSTARRS).  As it is a single object, it behooves us to consider whether we have overlooked some bias that has caused us to overinterpret an unusual object.  Certainly, unique or unexpected objects will attract attention from astronomers; however as the authors had begun work on this problem before the discovery of C/2014 S3 (PANSTARRS) we do not believe this effect should be significant.  Ultimately, the Large Synoptic Survey Telescope should improve the measured rate of C/2014 S3 (PANSTARRS)-like object, either confirming our conclusions here, or finding that the discovery of C/2014 S3 (PANSTARRS) was an extremely unlikely event.

\section{Acknowledgements}
We thank the referee Ramon Brasser for a report that, in particular, improved the clarity and context of this manuscript.  AS is partially supported by funding from the Center for Exoplanets and Habitable Worlds. The Center for Exoplanets and Habitable Worlds is supported by the Pennsylvania State University, the Eberly College of Science, and the Pennsylvania Space Grant Consortium.  AS and MW acknowledge support by the European Union through ERC grant number 279973.

{\footnotesize
\bibliographystyle{mnras}
\bibliography{oortcloudasteroids}
}

\label{lastpage}
\end{document}